\documentclass[aps,prl,twocolumn,notitlepage,superscriptaddress]{revtex4-1}  
\usepackage[textwidth=18cm,textheight=25cm]{geometry}
\usepackage[utf8]{inputenc}
\usepackage{color}
\usepackage{scalerel}

\usepackage{amsmath}
\usepackage{amssymb} 
\usepackage{mathtools}
\usepackage[hidelinks,breaklinks=true]{hyperref}
\usepackage{float}
\usepackage{physics}

\usepackage[symbol]{footmisc}

\begin{document}
\date{\today}

\title{Local dynamical heterogeneity in simple glass formers}

\author{Giulio Biroli}
\affiliation{Laboratoire de Physique de l'Ecole Normale Sup\'erieure, ENS, Universit\'e PSL, CNRS, Sorbonne Universit\'e, Universit\'e de Paris, F-75005 Paris, France
}
\author{Patrick Charbonneau}
\affiliation{Department of Chemistry, Duke University, Durham, North Carolina 27708, USA}
\affiliation{Department of Physics, Duke University, Durham, North Carolina 27708, USA}
\author{Giampaolo Folena}
\affiliation{Laboratoire de Physique de l'Ecole Normale Sup\'erieure, ENS, Universit\'e PSL, CNRS, Sorbonne Universit\'e, Universit\'e de Paris, F-75005 Paris, France
}
\affiliation{James Franck Institute and Department of Physics, University of Chicago, Chicago, Illinois 60637, USA}
\author{Yi Hu}
\affiliation{Department of Chemistry, Duke University, Durham, North Carolina 27708, USA}
\author{Francesco Zamponi}
\affiliation{Laboratoire de Physique de l'Ecole Normale Sup\'erieure, ENS, Universit\'e PSL, CNRS, Sorbonne Universit\'e, Universit\'e de Paris, F-75005 Paris, France
}

\begin{abstract}
We study the local dynamical fluctuations in glass-forming models of particles embedded in $d$-dimensional space, in the mean-field limit of $d\to\infty$. Our analytical calculation reveals that single-particle observables, such as squared particle displacements, display divergent fluctuations around the dynamical (or mode-coupling) transition, due to the emergence of nontrivial correlations between displacements along different directions. This effect notably gives rise to a divergent non-Gaussian parameter, $\alpha_2$. The $d\to\infty$ local dynamics therefore becomes quite rich upon approaching the glass transition. The finite-$d$ remnant of this phenomenon further provides a long sought-after, first-principle explanation for the growth of $\alpha_2$ around the glass transition that is \emph{not based on multi-particle correlations}. 
\end{abstract}
	
\maketitle

\paragraph{Introduction --}
Assessing the role of local order and of more extended static correlations 
on the dynamics of deeply supercooled liquids is one of the foremost open problems in the physics of glasses.
First-principle descriptions give little weight to either. The mode-coupling theory of glasses (MCT) proposes that a self-consistent freezing of density fluctuations leads to particle caging~\cite{gotze2009}, and hence that neither local order nor high-order correlations play a role in glass formation. Similarly, the mean-field theory of glasses, which is exactly realized for $d$-dimensional particles in the limit $d\rightarrow\infty$~\cite{KW87}, 
reduces the problem to a self-consistent description of binary collisions between particle pairs~\cite{maimbourg2016solution,francesco2020theory,liu2021dynamics}. By contrast, numerical simulations and colloidal experiments have identified a relatively strong correlation between (local) structure and dynamical fluctuations (see e.g.~\cite{widmer-cooper2004,royall2015,schoenholz2016}). The investigation of these dynamical correlations between particles has further uncovered a strong dynamical heterogeneity~\cite{schmidt1991nature,ediger2000spatially,BBBCS11}, as manifested by a strong spatial correlation between particle displacements over a dynamical length scale $\xi_4$ associated to four-point spatio-temporal correlations.
Standard MCT~\cite{BBMR06} and $d\to\infty$ theories~\cite{KT87,KT87b,FP00,franz2011field,franz2013static} find these correlations to be associated with \emph{collective excitations}, leading to a divergent $\xi_4$, and thus to a diverging dynamical susceptibility $\chi_4$ (from integrating the correlations over $\xi_4$) at the glass transition~\cite{berthier2011theoretical}.

The extent to which the complex glassy dynamics observed in bulk correlation functions is due to the superposition of many simple yet heterogeneous local relaxations, or to an inherently complex particle-level dynamics nevertheless remains debated~\cite{ediger2000spatially}. The riddle is vividly illustrated by the strong growth around the glass transition of a simple single-particle observable that quantifies the non-Gaussian character of displacement fluctuations, the long-studied non-Gaussian parameter $\alpha_2(t)$~\cite{kob1997dynamical}.
A possible explanation for this growth is that in the supercooled regime, at intermediate times different populations of particles --fast and slow~\cite{kob1997dynamical}-- emerge; both display normal diffusion but their superposed behavior leads to an apparent non-Gaussianity~\cite{chaudhuri2009diffusion}. Yet the remarkable similarity between both the time evolution and the temperature dependence of $\alpha_2$ and those of $\chi_4$ suggest that growing particle-level fluctuations might then also play a role. Unlike $\chi_4$, however, $\alpha_2$ cannot capture spatial correlations between displacements of distinct particles. In fact, generally speaking, single-particle observables cannot account for divergent spatial correlations around phase transitions, which are collective in nature.
The long-puzzling question is thus: why does the dynamics of a single particle display growing fluctuations around
the glass transition? Are these fluctuations a signature of a complex {\it local} dynamics? Or can this phenomenology be explained in terms of \emph{collective} dynamical heterogeneity? An early MCT study of $\alpha_2(t)$ supports the former interpretation~\cite{fuchs1998asymptotic}, but the poor quantitative performance and the approximate nature of the treatment left the matter unsettled.

Recently, it was shown that the \emph{single-particle} random Lorentz gas (RLG) model---a point tracer navigating within Poisson distributed spherical obstacles---shares key mean-field physics with structural glasses. Although the tracer dynamics is known to follow the percolation physics in $d=2,3$~\cite{hofling2006localization,hofling2008critical,bauer2010localization}, it presents the same caging physics as a many-body hard sphere (HS) liquid in the limit $d\rightarrow\infty$. Finite-$d$ corrections evaluated from numerical simulations were further found to seemingly diverge around the (avoided) dynamical transition~\cite{biroli2021mean,biroli2021unifying}.
Because standard explanations in terms of collective effects cannot be invoked--given that these effects are by construction absent from the RLG--these observations are particularly confounding. Teasing out the underlying microscopic mechanisms could thus illuminate some of the key conundrums of glass physics.

In this Letter, we investigate analytically and numerically the finite-$d$ perturbative fluctuations around the $d\rightarrow\infty$ solution of both HS liquids and the RLG. We define the appropriate dynamical susceptibility that probes fluctuations of the local dynamics, $\chi_4^\mathrm{S}$, and show that this \emph{single-particle} susceptibility diverges upon approaching the dynamical (or mode-coupling) transition. Our analytical expressions for the components of $\chi_4^\mathrm{S}$ also robustly and quantitatively match numerical results obtained up to $d=20$. The physical origin of these divergent fluctuations is traced back to the emergence of nontrivial correlations between single-particle displacements along distinct spatial directions. (In the limit $d\to\infty$, a sufficient number of distinct spatial directions exist for $\chi_4^\mathrm{S}$ to diverge.)
Because $\chi_4^\mathrm{S}$ is directly proportional to the non-Gaussian parameter $\alpha_2$, the latter
then also diverges. We further show numerically that the behavior of $\alpha_2$ in the diffusive phase of the RLG is qualitatively similar to that of three-dimensional glass-forming liquids. Our results thus reveal that in large dimension the dynamics of supercooled liquids becomes inherently complex {\it even at the local level}, and that the $\chi_4$-like behavior of $\alpha_2$ is a signature of this phenomenon.  The growth of $\alpha_2$ in three-dimensional supercooled liquids can then be naturally explained as a remnant of the avoided criticality of the dynamical transition.

\paragraph{Measures of dynamical fluctuations --} For simplicity, we henceforth mostly focus on the RLG, but our results can be straightforwardly generalized to standard HS and other simple glass formers (in the sense of Ref.~\cite{francesco2020theory}), which behave similarly locally in the limit $d\rightarrow\infty$. Their descriptions are then indeed equivalent after trivially rescaling the scaled packing fraction (of obstacles), $\hat\varphi = 2^d \varphi/d$, where $\varphi$ is the standard packing fraction (see Ref.~\cite{biroli2021unifying} for details). Advantageously, numerical simulations of the RLG can make use of ``quiet planting''~\cite{krzakala2009hiding,coja2013quiet,CJPZ14}, with each system realization having a tracer at the origin at time $t=0$ and obstacles drawn at random uniformly yet compatibly with the tracer position~\cite{biroli2021mean}. Newtonian dynamics can then be used to follow the tracer displacement, $\vec r(t) = \{r_1(t),\cdots,r_d(t)\}$, with the initial velocity being chosen uniformly at random with unit modulus. 

The standard non-Gaussian parameter can be generalized to $d$ spatial dimensions as~\cite{huang2015non}
\begin{equation} \label{eq:nongaussian}
\alpha_2(t) = \frac{d}{d+2}  \frac{ [\langle r^4 (t) \rangle]}{[ \langle r^2 (t) \rangle]^2 } - 1 \ ,
\end{equation}
where $\langle \cdots \rangle$ denotes dynamical (thermal) averaging of observables for a given set of obstacles, and $[ \cdots ]$ denotes averaging over different obstacle positions (quenched disorder). (This double averaging is equivalent to averaging simultaneously over dynamical trajectories and particles in supercooled liquids~\cite{kob1997dynamical}.)
We then introduce the relative variance (or kurtosis) and covariance of the spatial displacements of the tracer associated to different spatial directions of the orthonormal frame,
\begin{equation}
\mathrm{K}(t) = 
\frac{[\langle r^4_i(t) \rangle]}{[\langle r^2_i(t) \rangle]^2} \ ,
\,
\mathrm{C}(t) = 
\frac{[\langle r^2_i(t) r^2_j(t)\rangle]}{[\langle r^2_i(t) \rangle]^2}-1 = \frac{\hat{\mathrm{C}}(t)}d \ .
\end{equation} 
By isotropy of space, $\mathrm{K}(t)$ and $\mathrm{C}(t)$ do not depend on the choice of indices (for $i\neq j$). We can therefore write 
\begin{equation}\label{eq:A}
[\langle r^4 (t) \rangle] = [\langle r^2_i(t) \rangle]^2 \{ d \, \mathrm{K}(t)  + d(d-1) (1+\mathrm{C}(t))\} \ .
\end{equation}
If the components $r_i(t)$ are independent and Gaussian distributed, then $ \mathrm{K}(t)=3$ and $\mathrm{C}(t) =0$, and hence $\alpha_2(t)=0$.  A non-zero $\alpha_2(t)$ thus indicates either a non-Gaussian distribution of $r_i(t)$, or a non-zero correlation between $r_i(t)$ and $r_j(t)$ with $i\neq j$.

\begin{figure}[t]
\centering
\includegraphics[width=0.5\textwidth]{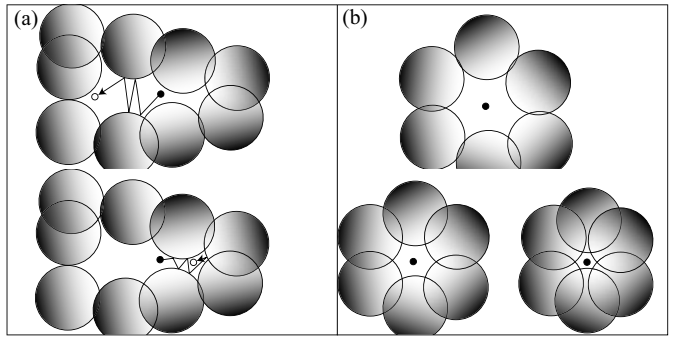}
\caption{Schematic representation of the two components of $\chi_4^{S}$: (a) thermal displacement fluctuations within a given cage $\chi^{S}_\mathrm{th}$; and (b) heterogeneous displacement fluctuations between different cages $\chi^{S}_\mathrm{het}$.}
\label{fig:illus}
\end{figure}

In the large $d$ limit, components of the displacement 
are known to scale as $r_i \sim 1/d$~\cite{francesco2020theory}. The mean squared displacement (MSD), which sums over these (squared) components, is therefore also of order $1/d$. It is then convenient to define a scaled MSD~\cite{francesco2020theory},
\begin{equation}
\hat\Delta(t) = d \Delta(t) = d  [\langle r(t)^2 \rangle] \ ,
\end{equation}
which remains finite as $d\to\infty$. Similarly, according to the central limit theorem (assuming that displacement along different directions are not too strongly correlated), its variance should scale as $\text{Var}(\hat\Delta)\sim 1/d$. From these quantities, we introduce local analogs to the standard $\chi_4$ for supercooled liquids~\cite{BBBCS11,franz2011field,franz2013static} (Fig.~\ref{fig:illus}).
The {\it single-particle} heterogeneous susceptibility is the variance of the thermally-averaged squared displacement over different cages,
\begin{equation}
\chi^\mathrm{S}_\mathrm{het}(t) = [\langle  r^2(t) \rangle^2] -  [\langle r^2(t) \rangle]^2 \ ,
\end{equation}
whereas the {\it single-particle} thermal susceptibility is the variance of thermal fluctuations evaluated for each cage and then averaged over all cages,
\begin{equation}
\chi^\mathrm{S}_\mathrm{th}(t) = [\langle r^4(t) \rangle -  \langle  r^2(t)\rangle^2] \ .
\end{equation}
The total {\it single-particle} susceptibility then reads
$\chi_4^\mathrm{S}(t) = \chi^\mathrm{S}_\mathrm{het}(t) + \chi^\mathrm{S}_\mathrm{th}(t)$. (Because these susceptibilities are all related to $\text{Var}(\Delta)\sim \text{Var}(\hat\Delta)/d^2\sim 1/d^3$, we define $\hat{\chi}^\mathrm{S}_{\mathrm{x}}=\chi^\mathrm{S}_{\mathrm{x}} \, d^3$, which remain finite as $d\to\infty$.)
The total susceptibility is further related to the non-Gaussian parameter as
\begin{equation} \label{eq:chi4scaled}
\frac{ \hat\chi_4^\mathrm{S}(t)}{d \hat\Delta(t)^2} = \frac{ [\langle r^4 (t) \rangle]}{ [\langle r^2 (t) \rangle]^2 } - 1  = \frac{d+2}{d} \alpha_2(t) + \frac2d \ .
\end{equation}
For $d\rightarrow\infty$, $\hat\alpha_2(t)=d\alpha_2(t)$ is thus given by the relative fluctuations of the caging order parameter. In other words, a growing $\hat \alpha_2$ indicates anomalously large dynamical fluctuations relative to those of simple (low-density) systems. 

As we shall show below, $\hat \alpha_2$ diverges around the  $d\rightarrow\infty$ dynamical transition. To pinpoint the underlying physical explanation, note that combining Eqs.~\eqref{eq:A} and \eqref{eq:chi4scaled} gives 
\begin{equation} 
\hat \alpha_2(t)= \mathrm{K}(t)-3 + \hat{\mathrm{C}}(t) \ .
\end{equation}
Because away from the transition $\hat\chi_4^\mathrm{S}(t)/\hat\Delta^2$-- and hence $\hat \alpha_2(t)$--is finite for ${d\to\infty}$, we deduce that both $\mathrm{K}(t) $ and $\hat{\mathrm{C}}(t)$ are finite, i.e., that the covariance $\mathrm{C}(t) \sim 1/d$. 
It is physically expected that $\mathrm{K}(t)$ remains finite at the critical point. (Although a formal proof is beyond the scope of this work, it would likely be based on the cavity method developed in Ref.~\cite{liu2021dynamics}, which suggests that all moments of the probability distribution of a properly scaled single component $r_i(t)$ remain finite as $d\to\infty$.)
A divergent susceptibility is thus necessarily related to anomalously large correlations of distinct spatial components of the displacement, i.e., a diverging $\hat{\mathrm{C}}(t)$.

\begin{figure}[t]
\centering
\includegraphics[width=0.5\textwidth]{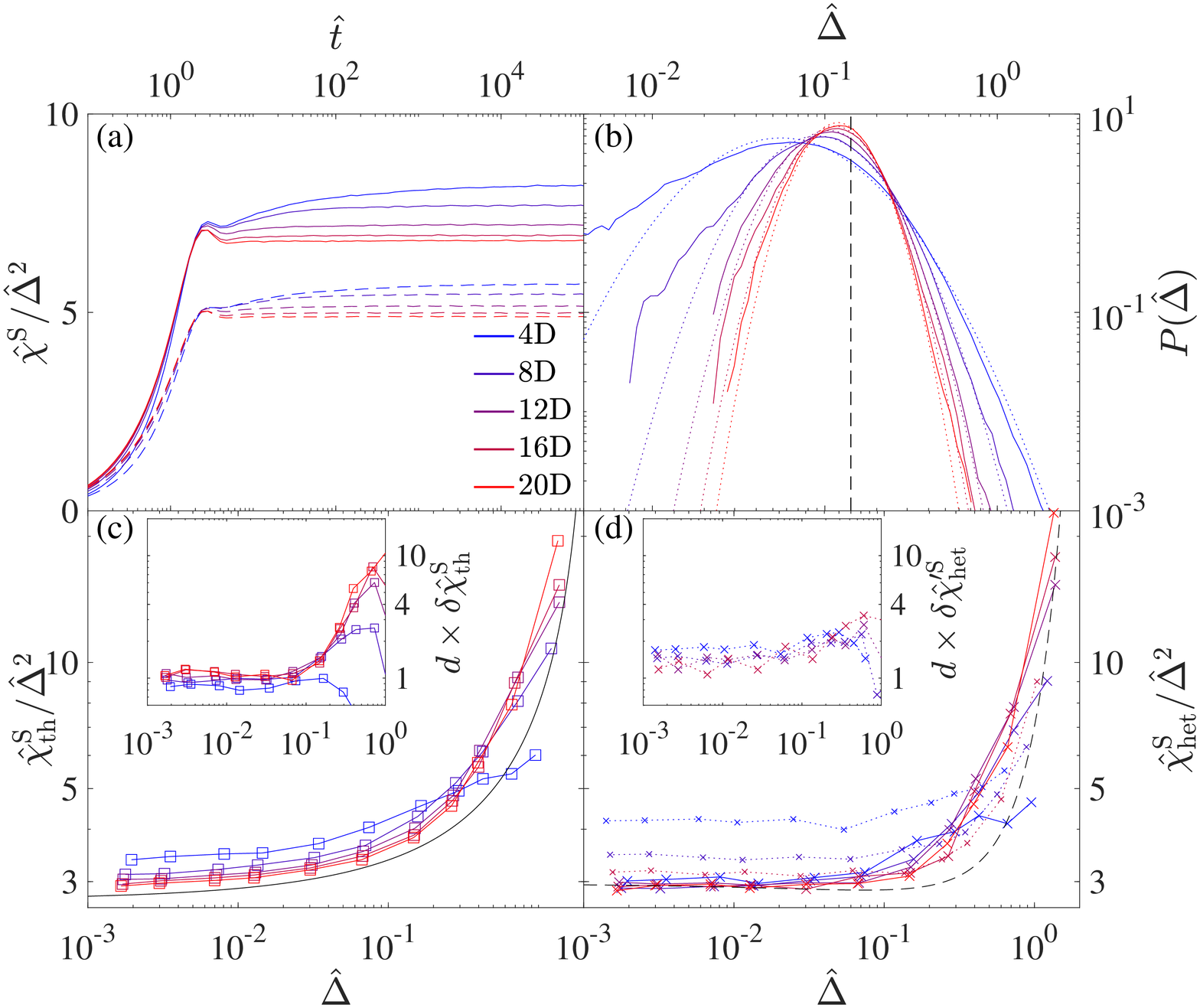}
\caption{Cage susceptibilities for the RLG. (a) Time evolution of $\hat\chi_4^\mathrm{S}$ (solid lines) and $\hat\chi^\mathrm{S}_\mathrm{th}$ (dashed lines) at $\hat\varphi=5$ for $d=4,8,12,16,20$ (blue to red), for scaled time $\hat{t}=t\sqrt{d} $, such that the MSD initially grows as $\hat{\Delta}(t) = \hat{t}^2$ in all dimensions~\cite{biroli2021unifying}.
(b) For the same systems, the distribution of the long-time MSD for different cages is well fitted by a log-normal form (dotted lines) around its maximum. The modal displacement steadily approaches the MFT prediction as $d$ increases (dashed line). The variance of the distribution provides an estimate of $\hat{\chi}^\mathrm{S}_\mathrm{het}$. 
(c) $\hat{\chi}^\mathrm{S}_\mathrm{th}$ and (d) $\hat{\chi}^\mathrm{S}_\mathrm{het}$ compared with $\hat{\chi}^{\mathrm{MFT}}$ (black lines) in $d=4$--$20$ (colors as in (a)). The log-normal fit in (b) provides an alternate estimate,  $\hat{\chi}'^{\mathrm{S}}_\mathrm{het}$ (dotted lines).  Insets of (c, d): The relative deviation of $\hat{\chi}^\mathrm{S}_\mathrm{th}$ and $\hat{\chi}'^{\mathrm{S}}_\mathrm{het}$ from the $d\rightarrow\infty$ prediction vanishes as $\sim 1/d$.}
\label{fig:chi}
\end{figure}

\paragraph{Results --} 
We first consider the high-density regime, in which numerical results can be directly compared with analytical predictions obtained from mean-field theory (MFT). For the glass phase, $\hat\varphi > \hat\varphi_d \approx 2.4$ in ${d\to\infty}$~\cite{biroli2021mean,biroli2021unifying}, the replica method detailed in Ref.~\onlinecite{francesco2020theory} provides the long-time limit of the MSD, ${\hat{\chi}^\mathrm{MFT}(t \rightarrow \infty) \equiv \hat{\chi}^\mathrm{MFT}}$. Based on the approach of Refs.~\onlinecite{franz2011field,franz2013static}, we obtain analytical expressions for $\hat\chi_\mathrm{th}^\mathrm{MFT}$ and $\hat\chi_\mathrm{het}^\mathrm{MFT}$, which will be published separately~\cite{bcfhz2021long}. These results show that both $\hat\chi_\mathrm{th}^\mathrm{MFT}$ and $\hat\chi_\mathrm{het}^\mathrm{MFT}$ are finite far from the dynamical transition, and diverge as $(\hat\varphi- \hat\varphi_d)^{-1/2}$ and $(\hat\varphi- \hat\varphi_d)^{-1}$, respectively, upon approaching that transition from the glass phase. 
Numerical results for $d\leq 20$, obtained as in Refs.~\cite{biroli2021mean,biroli2021unifying}, are consistent with these findings. Figure~\ref{fig:chi}a shows the saturation of the different susceptibilities at long times, whereas Fig.~\ref{fig:chi}b displays the distribution of MSD over different cages. The finite-dimensional results for both $\hat\chi^\mathrm{S}_\mathrm{th}$ and $\hat\chi^\mathrm{S}_\mathrm{het}$ converge to the mean-field predictions upon increasing $d$ (Fig.~\ref{fig:chi}c, d), and increase dramatically upon approaching $\hat\varphi_\mathrm{d}$ (or increasing $\hat \Delta$). 
This growth explains the diverging dimensional correction prefactor to the cage sizes reported in Ref.~\onlinecite{biroli2021unifying}.

Finite-$d$ corrections to these susceptibilities provide additional physical insights. For $\hat\chi^\mathrm{S}_\mathrm{th}$, in particular, 
numerical results clearly hint at a perturbative $1/d$ correction to the mean-field calculation, related to an even higher-order susceptibility that also diverges at $\hat\varphi_\mathrm{d}$.
However, in the vicinity of the dynamical transition both $\hat\chi^\mathrm{S}_\mathrm{th}$ and $\hat\chi^\mathrm{S}_\mathrm{het}$ deviate more strongly from the analytical prediction. This behavior is akin to the anomalous scaling of the mean cage size, understood in Ref.~\onlinecite{biroli2021unifying} to arise from rare large cages and other fluctuations~\cite{rizzo2020solvable} that are neglected by the perturbative calculation. To screen out these contributions, we proposed in Ref.~\onlinecite{biroli2021unifying} to consider the modal instead of the mean cage size. We can here similarly fit the cage size distribution around its mode with a log-normal form, and obtain from its variance an alternative estimate, ${\chi'}^\mathrm{S}_\mathrm{het}$ (Fig.~\ref{fig:chi}b). The resulting anomalies close to $\hat\varphi_\mathrm{d}$ are then somewhat reduced  (Fig.~\ref{fig:chi}d), and the $1/d$ correction is then recovered as well, albeit with a relatively large prefactor at small $\hat\Delta$. 

\begin{figure}[t]
\centering
\includegraphics[width=0.5\textwidth]{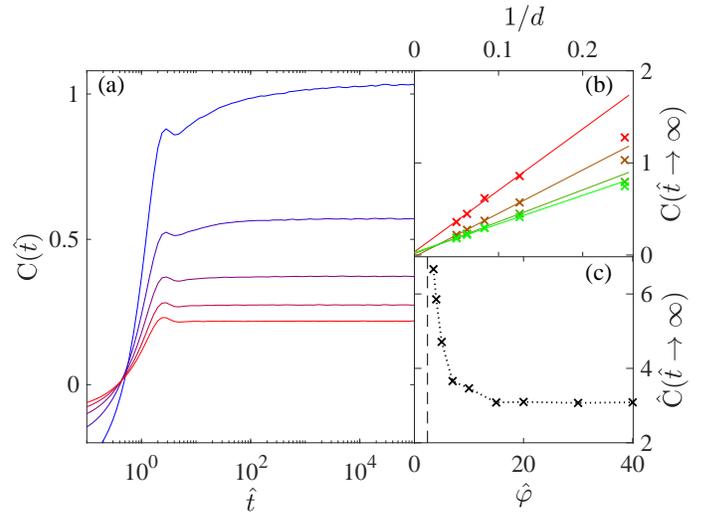}
\caption{Covariance $\mathrm{C}(t)$ between distinct squared coordinates of the RLG dynamics. (a) Time evolution of $\mathrm{C}(t)$ for $\hat\varphi=5$ and $d=4,8,12,16,20$, from blue to red. (b) Perturbative $1/d$ scaling of the long-time limit of $\mathrm{C}$ for $\hat\varphi=3.5,5,10,20$, from red to green. Lines are fits for $d \ge 8$. (c) The rescaled $\hat{\mathrm{C}}$ obtained from the fitted slopes in (b) grow markedly upon approaching $\hat\varphi_\mathrm{d}$ (the vertical dashed line) from above.}
\label{fig:ncov}
\end{figure}

To validate the physical interpretation of the divergence of the single-particles susceptibilities, we explicitly consider correlations between displacements along distinct spatial directions. The covariance $\mathrm{C}(t)$ indeed plateaus at long times in the glass phase (Fig.~\ref{fig:ncov}a), and that plateau, $\mathrm{C}= \mathrm{C}(t \rightarrow \infty)$, scales as $1/d$ (Fig.~\ref{fig:ncov}b).
Its rescaled counterpart $\hat{\mathrm{C}} = d\mathrm{C}$ is found to be almost constant at large $\hat\varphi \ge 15$, but increases significantly upon approaching $\hat{\varphi}_\mathrm{d}$ (Fig.~\ref{fig:ncov}c), similarly to $\hat\chi_\mathrm{th}$. The local dynamics then does become strongly correlated along distinct spatial directions, which suggests that cage shapes are increasingly complex. As a result, large coordinate displacements in correlated directions are required for their exploration.

\begin{figure}[t]
\centering
\includegraphics[width=0.3\textwidth]{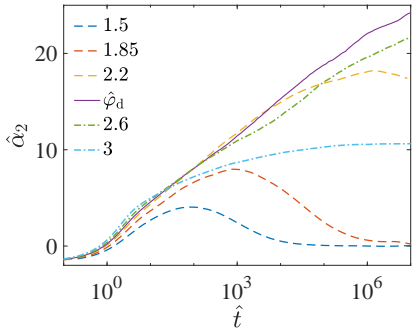}
\caption{Time evolution of the non-Gaussian parameter $\hat{\alpha}_2$ for different $\hat\varphi$ in $d=8$. The peak and plateau heights grow as $\hat\varphi \rightarrow \hat\varphi_\mathrm{d}$ from below (dashed lines) and from above (dot-dashed lines), respectively.}
\label{fig:a2}
\end{figure}

We complement this analysis with a consideration of $\alpha_2(t)$ at low packing fractions, for which only numerical simulations results are available. Developing an analytical approach would require combining the numerical solution of dynamical mean-field theory~\cite{manacorda2020numerical} with the methods of Refs.~\cite{franz2011field,franz2013static}---a very challenging task. We focus on large-dimensional systems (Fig.~\ref{fig:a2}), since we have established in Refs.~\cite{biroli2021mean,biroli2021unifying} that for $d\gtrsim 8$ the dynamical arrest of the RLG takes place at $\hat\varphi > \hat\varphi_\mathrm{d}$~\cite{footnote}. (While for $d\lesssim 8$ percolation physics controls the low-density behavior of this particular model, because the percolation threshold $\hat{\varphi}_\mathrm{p} < \hat{\varphi}_\mathrm{d}$, the tracer is already localized within percolation cage as $\hat{\varphi}$ approach $\hat{\varphi}_\mathrm{d}$ from below and follows the percolation criticality~\cite{charbonneau2021percolation}.) At low density, $\alpha_2(t)$ peaks at intermediate times and then fully decays at long times, as a result of the tracer diffusing. Upon increasing density, the peak grows as does its associated timescale in a fashion strongly reminiscent of three-dimensional supercooled liquids~\cite{kob1997dynamical}.
At high densities ---beyond the dynamical glass transition--- $\alpha_2(t\to\infty)$ instead plateaus because caging persist at long times (as analyzed in Fig.~\ref{fig:chi}). The divergence of the plateau height for $\hat\varphi \rightarrow \hat \varphi_d$ from above is thus consistent with a divergence of the peak height for $\hat\varphi \rightarrow \hat \varphi_d$ from below. They are two facets of the same phenomenon. The {\it local} glassy dynamics thus becomes increasingly complex and correlated in the vicinity of the dynamical transition.

\paragraph{Conclusions --} We have here specifically explored the perturbative finite-$d$ corrections to the random Lorentz gas by computing analytically and numerically the fluctuations of the caging order parameter in the dynamically arrested phase. In this single-particle model, when $d$ is large enough, dynamical heterogeneity of a peculiar kind gives rise to a divergent single particle susceptibility, $\chi^\mathrm{S}_4$, as $d\rightarrow\infty$. Even in the absence of any many-body effect, spatial components  of the tracer displacement are then sufficiently numerous for their dynamical correlations to diverge. Correspondingly, the single-particle non-Gaussian parameter $\alpha_2$ then also displays a divergent peak, both at long times, upon approaching the transition from the glass phase, and at intermediate times, upon approaching the transition from the liquid phase. 
Because when $d\to\infty$ the RLG model is completely equivalent to HS and other simple glass formers, our results also apply to these systems. (For $d = 3$ liquids, however, activated processes might obfuscate this prediction, especially in the close vicinity of $\hat{\varphi}_\mathrm{d}$.) In the infinite dimensional limit, the many-body dynamical slowing down is thus accompanied by two kinds of diverging fluctuations: those associated with collective dynamical heterogeneity and those revealing an increasingly complex local dynamics. A thorough study of dynamical heterogeneity in large $d$ for the HS model (or other many-body models) is left as future work, but by analogy with other infinite-dimensional models we expect correlations between close neighbors to then play a key role~\cite[Chap.~1]{francesco2020theory}. 

Although the local dynamical fluctuations analyzed in this work only diverge in the infinite-dimensional limit, their finite-dimensional echo putatively explains the increase of $\alpha_2(t)$, and its similarity with $\chi_4$, in three-dimensional glass formers. A similar phenomenon is also expected in the vicinity of other phase transitions, such as the Gardner transition~\cite{francesco2020theory,BCJPSZ15}.
Our results thus offer a first-principle resolution to the puzzling observation that supercooled liquids display significant complex and correlated dynamics at the single-particle level. 
The associated non-monotonicity prediction should be directly testable by numerical simulations of polydisperse HS, which can be studied across a wide range of time scales and spatial dimensions~\cite{berthier2019finite}. Because both contributions to $\chi_4^{S}$ encode geometrical information, this prediction also offers a first-principle explanation for at least part of the role played by cage structure on single-particle dynamics~\cite{royall2015}.

A natural next step would be to consider non-perturbative fluctuations via instantonic dynamical calculations~\cite{rizzo2021path}. Such corrections are ultimately responsible for the dynamical transition being avoided in finite $d$, which in the RLG happens via single-particle hopping~\cite{CJPZ14,biroli2021mean} and in glass formers via both single-particle hopping and many-body collective effects such as facilitation and nucleation~\cite{KTW89,KHGGC11,guiselin2021microscopic}. The RLG could thus prove particularly useful for examining single-particle non-perturbative contributions, as a first leap towards understanding more complicated collective effects.

\paragraph{Acknowledgments --}
We thank Ludovic Berthier, Silvio Franz and Federico Ricci-Tersenghi for helpful discussions. This project has received funding from the European Research Council (ERC) under the European Union's Horizon 2020 research and innovation programme (grant agreement n. 723955 - GlassUniversality) and from the Simons Foundation (Grant No. 454935 to G.B.; Grant No. 454937 to P.C.;  Grant No. 454955 to F.Z.).
 
\bibliography{./bibliography.bib}

\begin{thebibliography}{41}%
\makeatletter
\providecommand \@ifxundefined [1]{%
 \@ifx{#1\undefined}
}%
\providecommand \@ifnum [1]{%
 \ifnum #1\expandafter \@firstoftwo
 \else \expandafter \@secondoftwo
 \fi
}%
\providecommand \@ifx [1]{%
 \ifx #1\expandafter \@firstoftwo
 \else \expandafter \@secondoftwo
 \fi
}%
\providecommand \natexlab [1]{#1}%
\providecommand \enquote  [1]{``#1''}%
\providecommand \bibnamefont  [1]{#1}%
\providecommand \bibfnamefont [1]{#1}%
\providecommand \citenamefont [1]{#1}%
\providecommand \href@noop [0]{\@secondoftwo}%
\providecommand \href [0]{\begingroup \@sanitize@url \@href}%
\providecommand \@href[1]{\@@startlink{#1}\@@href}%
\providecommand \@@href[1]{\endgroup#1\@@endlink}%
\providecommand \@sanitize@url [0]{\catcode `\\12\catcode `\$12\catcode
  `\&12\catcode `\#12\catcode `\^12\catcode `\_12\catcode `\%12\relax}%
\providecommand \@@startlink[1]{}%
\providecommand \@@endlink[0]{}%
\providecommand \url  [0]{\begingroup\@sanitize@url \@url }%
\providecommand \@url [1]{\endgroup\@href {#1}{\urlprefix }}%
\providecommand \urlprefix  [0]{URL }%
\providecommand \Eprint [0]{\href }%
\providecommand \doibase [0]{http://dx.doi.org/}%
\providecommand \selectlanguage [0]{\@gobble}%
\providecommand \bibinfo  [0]{\@secondoftwo}%
\providecommand \bibfield  [0]{\@secondoftwo}%
\providecommand \translation [1]{[#1]}%
\providecommand \BibitemOpen [0]{}%
\providecommand \bibitemStop [0]{}%
\providecommand \bibitemNoStop [0]{.\EOS\space}%
\providecommand \EOS [0]{\spacefactor3000\relax}%
\providecommand \BibitemShut  [1]{\csname bibitem#1\endcsname}%
\let\auto@bib@innerbib\@empty
\bibitem [{\citenamefont {G\"otze}(2009)}]{gotze2009}%
  \BibitemOpen
  \bibfield  {author} {\bibinfo {author} {\bibfnamefont {W.}~\bibnamefont
  {G\"otze}},\ }\href {\doibase 10.1093/acprof:oso/9780199235346.001.0001}
  {\emph {\bibinfo {title} {Complex Dynamics of Glass-Forming Liquids}}},\
  \bibinfo {series} {International Series of Monographs on Physics}, Vol.\
  \bibinfo {volume} {143}\ (\bibinfo  {publisher} {Oxford University Press},\
  \bibinfo {address} {Oxford},\ \bibinfo {year} {2009})\ p.\ \bibinfo {pages}
  {641}\BibitemShut {NoStop}%
\bibitem [{\citenamefont {Kirkpatrick}\ and\ \citenamefont
  {Wolynes}(1987)}]{KW87}%
  \BibitemOpen
  \bibfield  {author} {\bibinfo {author} {\bibfnamefont {T.~R.}\ \bibnamefont
  {Kirkpatrick}}\ and\ \bibinfo {author} {\bibfnamefont {P.~G.}\ \bibnamefont
  {Wolynes}},\ }\href {\doibase 10.1103/PhysRevA.35.3072} {\bibfield  {journal}
  {\bibinfo  {journal} {Physical Review A}\ }\textbf {\bibinfo {volume} {35}},\
  \bibinfo {pages} {3072} (\bibinfo {year} {1987})}\BibitemShut {NoStop}%
\bibitem [{\citenamefont {Maimbourg}\ \emph {et~al.}(2016)\citenamefont
  {Maimbourg}, \citenamefont {Kurchan},\ and\ \citenamefont
  {Zamponi}}]{maimbourg2016solution}%
  \BibitemOpen
  \bibfield  {author} {\bibinfo {author} {\bibfnamefont {T.}~\bibnamefont
  {Maimbourg}}, \bibinfo {author} {\bibfnamefont {J.}~\bibnamefont {Kurchan}},
  \ and\ \bibinfo {author} {\bibfnamefont {F.}~\bibnamefont {Zamponi}},\ }\href
  {\doibase 10.1103/PhysRevLett.116.015902} {\bibfield  {journal} {\bibinfo
  {journal} {Phys. Rev. Lett.}\ }\textbf {\bibinfo {volume} {116}},\ \bibinfo
  {pages} {015902} (\bibinfo {year} {2016})}\BibitemShut {NoStop}%
\bibitem [{\citenamefont {Parisi}\ \emph {et~al.}(2020)\citenamefont {Parisi},
  \citenamefont {Urbani},\ and\ \citenamefont {Zamponi}}]{francesco2020theory}%
  \BibitemOpen
  \bibfield  {author} {\bibinfo {author} {\bibfnamefont {G.}~\bibnamefont
  {Parisi}}, \bibinfo {author} {\bibfnamefont {P.}~\bibnamefont {Urbani}}, \
  and\ \bibinfo {author} {\bibfnamefont {F.}~\bibnamefont {Zamponi}},\ }\href
  {\doibase 10.1017/9781108120494} {\emph {\bibinfo {title} {Theory of simple
  glasses: Exact Solutions in Infinite Dimensions}}}\ (\bibinfo  {publisher}
  {Cambridge University Press},\ \bibinfo {address} {Cambridge CB2 8BS, United
  Kingdom},\ \bibinfo {year} {2020})\BibitemShut {NoStop}%
\bibitem [{\citenamefont {Liu}\ \emph {et~al.}(2021)\citenamefont {Liu},
  \citenamefont {Biroli}, \citenamefont {Reichman},\ and\ \citenamefont
  {Szamel}}]{liu2021dynamics}%
  \BibitemOpen
  \bibfield  {author} {\bibinfo {author} {\bibfnamefont {C.}~\bibnamefont
  {Liu}}, \bibinfo {author} {\bibfnamefont {G.}~\bibnamefont {Biroli}},
  \bibinfo {author} {\bibfnamefont {D.~R.}\ \bibnamefont {Reichman}}, \ and\
  \bibinfo {author} {\bibfnamefont {G.}~\bibnamefont {Szamel}},\ }\href@noop {}
  {\bibfield  {journal} {\bibinfo  {journal} {Phys. Rev. E}\ }\textbf {\bibinfo
  {volume} {104}},\ \bibinfo {pages} {054606} (\bibinfo {year}
  {2021})}\BibitemShut {NoStop}%
\bibitem [{\citenamefont {Widmer-Cooper}\ \emph {et~al.}(2004)\citenamefont
  {Widmer-Cooper}, \citenamefont {Harrowell},\ and\ \citenamefont
  {Fynewever}}]{widmer-cooper2004}%
  \BibitemOpen
  \bibfield  {author} {\bibinfo {author} {\bibfnamefont {A.}~\bibnamefont
  {Widmer-Cooper}}, \bibinfo {author} {\bibfnamefont {P.}~\bibnamefont
  {Harrowell}}, \ and\ \bibinfo {author} {\bibfnamefont {H.}~\bibnamefont
  {Fynewever}},\ }\href {\doibase 10.1103/PhysRevLett.93.135701} {\bibfield
  {journal} {\bibinfo  {journal} {Phys. Rev. Lett.}\ }\textbf {\bibinfo
  {volume} {93}},\ \bibinfo {pages} {135701} (\bibinfo {year}
  {2004})}\BibitemShut {NoStop}%
\bibitem [{\citenamefont {Royall}\ and\ \citenamefont
  {Williams}(2015)}]{royall2015}%
  \BibitemOpen
  \bibfield  {author} {\bibinfo {author} {\bibfnamefont {C.~P.}\ \bibnamefont
  {Royall}}\ and\ \bibinfo {author} {\bibfnamefont {S.~R.}\ \bibnamefont
  {Williams}},\ }\href {\doibase https://doi.org/10.1016/j.physrep.2014.11.004}
  {\bibfield  {journal} {\bibinfo  {journal} {Phys. Rep.}\ }\textbf {\bibinfo
  {volume} {560}},\ \bibinfo {pages} {1} (\bibinfo {year} {2015})}\BibitemShut
  {NoStop}%
\bibitem [{\citenamefont {Schoenholz}\ \emph {et~al.}(2016)\citenamefont
  {Schoenholz}, \citenamefont {Cubuk}, \citenamefont {Sussman}, \citenamefont
  {Kaxiras},\ and\ \citenamefont {Liu}}]{schoenholz2016}%
  \BibitemOpen
  \bibfield  {author} {\bibinfo {author} {\bibfnamefont {S.~S.}\ \bibnamefont
  {Schoenholz}}, \bibinfo {author} {\bibfnamefont {E.~D.}\ \bibnamefont
  {Cubuk}}, \bibinfo {author} {\bibfnamefont {D.~M.}\ \bibnamefont {Sussman}},
  \bibinfo {author} {\bibfnamefont {E.}~\bibnamefont {Kaxiras}}, \ and\
  \bibinfo {author} {\bibfnamefont {A.~J.}\ \bibnamefont {Liu}},\ }\href
  {\doibase 10.1038/nphys3644} {\bibfield  {journal} {\bibinfo  {journal} {Nat.
  Phys.}\ }\textbf {\bibinfo {volume} {12}},\ \bibinfo {pages} {469} (\bibinfo
  {year} {2016})}\BibitemShut {NoStop}%
\bibitem [{\citenamefont {Schmidt-Rohr}\ and\ \citenamefont
  {Spiess}(1991)}]{schmidt1991nature}%
  \BibitemOpen
  \bibfield  {author} {\bibinfo {author} {\bibfnamefont {K.}~\bibnamefont
  {Schmidt-Rohr}}\ and\ \bibinfo {author} {\bibfnamefont {H.~W.}\ \bibnamefont
  {Spiess}},\ }\href {\doibase 10.1103/PhysRevLett.66.3020} {\bibfield
  {journal} {\bibinfo  {journal} {Phys. Rev. Lett.}\ }\textbf {\bibinfo
  {volume} {66}},\ \bibinfo {pages} {3020} (\bibinfo {year}
  {1991})}\BibitemShut {NoStop}%
\bibitem [{\citenamefont {Ediger}(2000)}]{ediger2000spatially}%
  \BibitemOpen
  \bibfield  {author} {\bibinfo {author} {\bibfnamefont {M.~D.}\ \bibnamefont
  {Ediger}},\ }\href {\doibase 10.1146/annurev.physchem.51.1.99} {\bibfield
  {journal} {\bibinfo  {journal} {Annu. Rev. Phys. Chem.}\ }\textbf {\bibinfo
  {volume} {51}},\ \bibinfo {pages} {99} (\bibinfo {year} {2000})}\BibitemShut
  {NoStop}%
\bibitem [{\citenamefont {Berthier}\ \emph {et~al.}(2011)\citenamefont
  {Berthier}, \citenamefont {Biroli}, \citenamefont {Bouchaud}, \citenamefont
  {Cipelletti},\ and\ \citenamefont {van Saarloos}}]{BBBCS11}%
  \BibitemOpen
  \bibfield  {author} {\bibinfo {author} {\bibfnamefont {L.}~\bibnamefont
  {Berthier}}, \bibinfo {author} {\bibfnamefont {G.}~\bibnamefont {Biroli}},
  \bibinfo {author} {\bibfnamefont {J.-P.}\ \bibnamefont {Bouchaud}}, \bibinfo
  {author} {\bibfnamefont {L.}~\bibnamefont {Cipelletti}}, \ and\ \bibinfo
  {author} {\bibfnamefont {W.}~\bibnamefont {van Saarloos}},\ }\href {\doibase
  10.1093/acprof:oso/9780199691470.001.0001} {\emph {\bibinfo {title}
  {Dynamical Heterogeneities and Glasses}}}\ (\bibinfo  {publisher} {Oxford
  University Press},\ \bibinfo {year} {2011})\BibitemShut {NoStop}%
\bibitem [{\citenamefont {Biroli}\ \emph {et~al.}(2006)\citenamefont {Biroli},
  \citenamefont {Bouchaud}, \citenamefont {Miyazaki},\ and\ \citenamefont
  {Reichman}}]{BBMR06}%
  \BibitemOpen
  \bibfield  {author} {\bibinfo {author} {\bibfnamefont {G.}~\bibnamefont
  {Biroli}}, \bibinfo {author} {\bibfnamefont {J.}~\bibnamefont {Bouchaud}},
  \bibinfo {author} {\bibfnamefont {K.}~\bibnamefont {Miyazaki}}, \ and\
  \bibinfo {author} {\bibfnamefont {D.}~\bibnamefont {Reichman}},\ }\href
  {\doibase 10.1103/PhysRevLett.97.195701} {\bibfield  {journal} {\bibinfo
  {journal} {Physical Review Letters}\ }\textbf {\bibinfo {volume} {97}},\
  \bibinfo {pages} {195701} (\bibinfo {year} {2006})}\BibitemShut {NoStop}%
\bibitem [{\citenamefont {Kirkpatrick}\ and\ \citenamefont
  {Thirumalai}(1987{\natexlab{a}})}]{KT87}%
  \BibitemOpen
  \bibfield  {author} {\bibinfo {author} {\bibfnamefont {T.~R.}\ \bibnamefont
  {Kirkpatrick}}\ and\ \bibinfo {author} {\bibfnamefont {D.}~\bibnamefont
  {Thirumalai}},\ }\href {\doibase 10.1103/PhysRevLett.58.2091} {\bibfield
  {journal} {\bibinfo  {journal} {Physical Review Letters}\ }\textbf {\bibinfo
  {volume} {58}},\ \bibinfo {pages} {2091} (\bibinfo {year}
  {1987}{\natexlab{a}})}\BibitemShut {NoStop}%
\bibitem [{\citenamefont {Kirkpatrick}\ and\ \citenamefont
  {Thirumalai}(1987{\natexlab{b}})}]{KT87b}%
  \BibitemOpen
  \bibfield  {author} {\bibinfo {author} {\bibfnamefont {T.~R.}\ \bibnamefont
  {Kirkpatrick}}\ and\ \bibinfo {author} {\bibfnamefont {D.}~\bibnamefont
  {Thirumalai}},\ }\href {\doibase 10.1103/PhysRevB.36.5388} {\bibfield
  {journal} {\bibinfo  {journal} {Physical Review B}\ }\textbf {\bibinfo
  {volume} {36}},\ \bibinfo {pages} {5388} (\bibinfo {year}
  {1987}{\natexlab{b}})}\BibitemShut {NoStop}%
\bibitem [{\citenamefont {Franz}\ and\ \citenamefont {Parisi}(2000)}]{FP00}%
  \BibitemOpen
  \bibfield  {author} {\bibinfo {author} {\bibfnamefont {S.}~\bibnamefont
  {Franz}}\ and\ \bibinfo {author} {\bibfnamefont {G.}~\bibnamefont {Parisi}},\
  }\href {\doibase 10.1088/0953-8984/12/29/305} {\bibfield  {journal} {\bibinfo
   {journal} {Journal of Physics: Condensed Matter}\ }\textbf {\bibinfo
  {volume} {12}},\ \bibinfo {pages} {6335} (\bibinfo {year}
  {2000})}\BibitemShut {NoStop}%
\bibitem [{\citenamefont {Franz}\ \emph {et~al.}(2011)\citenamefont {Franz},
  \citenamefont {Parisi}, \citenamefont {Ricci-Tersenghi},\ and\ \citenamefont
  {Rizzo}}]{franz2011field}%
  \BibitemOpen
  \bibfield  {author} {\bibinfo {author} {\bibfnamefont {S.}~\bibnamefont
  {Franz}}, \bibinfo {author} {\bibfnamefont {G.}~\bibnamefont {Parisi}},
  \bibinfo {author} {\bibfnamefont {F.}~\bibnamefont {Ricci-Tersenghi}}, \ and\
  \bibinfo {author} {\bibfnamefont {T.}~\bibnamefont {Rizzo}},\ }\href
  {\doibase 10.1140/epje/i2011-11102-0} {\bibfield  {journal} {\bibinfo
  {journal} {Eur. Phys. J. E}\ }\textbf {\bibinfo {volume} {34}},\ \bibinfo
  {pages} {102} (\bibinfo {year} {2011})}\BibitemShut {NoStop}%
\bibitem [{\citenamefont {Franz}\ \emph {et~al.}(2013)\citenamefont {Franz},
  \citenamefont {Jacquin}, \citenamefont {Parisi}, \citenamefont {Urbani},\
  and\ \citenamefont {Zamponi}}]{franz2013static}%
  \BibitemOpen
  \bibfield  {author} {\bibinfo {author} {\bibfnamefont {S.}~\bibnamefont
  {Franz}}, \bibinfo {author} {\bibfnamefont {H.}~\bibnamefont {Jacquin}},
  \bibinfo {author} {\bibfnamefont {G.}~\bibnamefont {Parisi}}, \bibinfo
  {author} {\bibfnamefont {P.}~\bibnamefont {Urbani}}, \ and\ \bibinfo {author}
  {\bibfnamefont {F.}~\bibnamefont {Zamponi}},\ }\href {\doibase
  10.1063/1.4776213} {\bibfield  {journal} {\bibinfo  {journal} {J. Chem.
  Phys.}\ }\textbf {\bibinfo {volume} {138}},\ \bibinfo {pages} {12A540}
  (\bibinfo {year} {2013})}\BibitemShut {NoStop}%
\bibitem [{\citenamefont {Berthier}\ and\ \citenamefont
  {Biroli}(2011)}]{berthier2011theoretical}%
  \BibitemOpen
  \bibfield  {author} {\bibinfo {author} {\bibfnamefont {L.}~\bibnamefont
  {Berthier}}\ and\ \bibinfo {author} {\bibfnamefont {G.}~\bibnamefont
  {Biroli}},\ }\href {\doibase 10.1103/RevModPhys.83.587} {\bibfield  {journal}
  {\bibinfo  {journal} {Rev. Mod. Phys.}\ }\textbf {\bibinfo {volume} {83}},\
  \bibinfo {pages} {587} (\bibinfo {year} {2011})}\BibitemShut {NoStop}%
\bibitem [{\citenamefont {Kob}\ \emph {et~al.}(1997)\citenamefont {Kob},
  \citenamefont {Donati}, \citenamefont {Plimpton}, \citenamefont {Poole},\
  and\ \citenamefont {Glotzer}}]{kob1997dynamical}%
  \BibitemOpen
  \bibfield  {author} {\bibinfo {author} {\bibfnamefont {W.}~\bibnamefont
  {Kob}}, \bibinfo {author} {\bibfnamefont {C.}~\bibnamefont {Donati}},
  \bibinfo {author} {\bibfnamefont {S.~J.}\ \bibnamefont {Plimpton}}, \bibinfo
  {author} {\bibfnamefont {P.~H.}\ \bibnamefont {Poole}}, \ and\ \bibinfo
  {author} {\bibfnamefont {S.~C.}\ \bibnamefont {Glotzer}},\ }\href {\doibase
  10.1103/PhysRevLett.79.2827} {\bibfield  {journal} {\bibinfo  {journal}
  {Phys. Rev. Lett.}\ }\textbf {\bibinfo {volume} {79}},\ \bibinfo {pages}
  {2827} (\bibinfo {year} {1997})}\BibitemShut {NoStop}%
\bibitem [{\citenamefont {Chaudhuri}\ \emph {et~al.}(2009)\citenamefont
  {Chaudhuri}, \citenamefont {Berthier}, \citenamefont {Sastry},\ and\
  \citenamefont {Kob}}]{chaudhuri2009diffusion}%
  \BibitemOpen
  \bibfield  {author} {\bibinfo {author} {\bibfnamefont {P.}~\bibnamefont
  {Chaudhuri}}, \bibinfo {author} {\bibfnamefont {L.}~\bibnamefont {Berthier}},
  \bibinfo {author} {\bibfnamefont {S.}~\bibnamefont {Sastry}}, \ and\ \bibinfo
  {author} {\bibfnamefont {W.}~\bibnamefont {Kob}},\ }\href@noop {} {\bibfield
  {journal} {\bibinfo  {journal} {diffusion-fundamentals.org}\ }\textbf
  {\bibinfo {volume} {11}},\ \bibinfo {pages} {55} (\bibinfo {year}
  {2009})}\BibitemShut {NoStop}%
\bibitem [{\citenamefont {Fuchs}\ \emph {et~al.}(1998)\citenamefont {Fuchs},
  \citenamefont {G\"otze},\ and\ \citenamefont {Mayr}}]{fuchs1998asymptotic}%
  \BibitemOpen
  \bibfield  {author} {\bibinfo {author} {\bibfnamefont {M.}~\bibnamefont
  {Fuchs}}, \bibinfo {author} {\bibfnamefont {W.}~\bibnamefont {G\"otze}}, \
  and\ \bibinfo {author} {\bibfnamefont {M.~R.}\ \bibnamefont {Mayr}},\ }\href
  {\doibase 10.1103/PhysRevE.58.3384} {\bibfield  {journal} {\bibinfo
  {journal} {Phys. Rev. E}\ }\textbf {\bibinfo {volume} {58}},\ \bibinfo
  {pages} {3384} (\bibinfo {year} {1998})}\BibitemShut {NoStop}%
\bibitem [{\citenamefont {H{\"o}fling}\ \emph {et~al.}(2006)\citenamefont
  {H{\"o}fling}, \citenamefont {Franosch},\ and\ \citenamefont
  {Frey}}]{hofling2006localization}%
  \BibitemOpen
  \bibfield  {author} {\bibinfo {author} {\bibfnamefont {F.}~\bibnamefont
  {H{\"o}fling}}, \bibinfo {author} {\bibfnamefont {T.}~\bibnamefont
  {Franosch}}, \ and\ \bibinfo {author} {\bibfnamefont {E.}~\bibnamefont
  {Frey}},\ }\href {\doibase 10.1103/PhysRevLett.96.165901} {\bibfield
  {journal} {\bibinfo  {journal} {Phys. Rev. Lett.}\ }\textbf {\bibinfo
  {volume} {96}},\ \bibinfo {pages} {165901} (\bibinfo {year}
  {2006})}\BibitemShut {NoStop}%
\bibitem [{\citenamefont {H{\"o}fling}\ \emph {et~al.}(2008)\citenamefont
  {H{\"o}fling}, \citenamefont {Munk}, \citenamefont {Frey},\ and\
  \citenamefont {Franosch}}]{hofling2008critical}%
  \BibitemOpen
  \bibfield  {author} {\bibinfo {author} {\bibfnamefont {F.}~\bibnamefont
  {H{\"o}fling}}, \bibinfo {author} {\bibfnamefont {T.}~\bibnamefont {Munk}},
  \bibinfo {author} {\bibfnamefont {E.}~\bibnamefont {Frey}}, \ and\ \bibinfo
  {author} {\bibfnamefont {T.}~\bibnamefont {Franosch}},\ }\href {\doibase
  10.1063/1.2901170} {\bibfield  {journal} {\bibinfo  {journal} {J. Chem.
  Phys.}\ }\textbf {\bibinfo {volume} {128}},\ \bibinfo {pages} {164517}
  (\bibinfo {year} {2008})}\BibitemShut {NoStop}%
\bibitem [{\citenamefont {Bauer}\ \emph {et~al.}(2010)\citenamefont {Bauer},
  \citenamefont {H{\"o}fling}, \citenamefont {Munk}, \citenamefont {Frey},\
  and\ \citenamefont {Franosch}}]{bauer2010localization}%
  \BibitemOpen
  \bibfield  {author} {\bibinfo {author} {\bibfnamefont {T.}~\bibnamefont
  {Bauer}}, \bibinfo {author} {\bibfnamefont {F.}~\bibnamefont {H{\"o}fling}},
  \bibinfo {author} {\bibfnamefont {T.}~\bibnamefont {Munk}}, \bibinfo {author}
  {\bibfnamefont {E.}~\bibnamefont {Frey}}, \ and\ \bibinfo {author}
  {\bibfnamefont {T.}~\bibnamefont {Franosch}},\ }\href {\doibase
  10.1140/epjst/e2010-01313-1} {\bibfield  {journal} {\bibinfo  {journal} {Eur.
  Phys. J Spec. Top.}\ }\textbf {\bibinfo {volume} {189}},\ \bibinfo {pages}
  {103} (\bibinfo {year} {2010})}\BibitemShut {NoStop}%
\bibitem [{\citenamefont {Biroli}\ \emph
  {et~al.}(2021{\natexlab{a}})\citenamefont {Biroli}, \citenamefont
  {Charbonneau}, \citenamefont {Hu}, \citenamefont {Ikeda}, \citenamefont
  {Szamel},\ and\ \citenamefont {Zamponi}}]{biroli2021mean}%
  \BibitemOpen
  \bibfield  {author} {\bibinfo {author} {\bibfnamefont {G.}~\bibnamefont
  {Biroli}}, \bibinfo {author} {\bibfnamefont {P.}~\bibnamefont {Charbonneau}},
  \bibinfo {author} {\bibfnamefont {Y.}~\bibnamefont {Hu}}, \bibinfo {author}
  {\bibfnamefont {H.}~\bibnamefont {Ikeda}}, \bibinfo {author} {\bibfnamefont
  {G.}~\bibnamefont {Szamel}}, \ and\ \bibinfo {author} {\bibfnamefont
  {F.}~\bibnamefont {Zamponi}},\ }\href {\doibase 10.1021/acs.jpcb.1c02067}
  {\bibfield  {journal} {\bibinfo  {journal} {J. Phys. Chem. B}\ }\textbf
  {\bibinfo {volume} {125}},\ \bibinfo {pages} {6244} (\bibinfo {year}
  {2021}{\natexlab{a}})}\BibitemShut {NoStop}%
\bibitem [{\citenamefont {Biroli}\ \emph
  {et~al.}(2021{\natexlab{b}})\citenamefont {Biroli}, \citenamefont
  {Charbonneau}, \citenamefont {Corwin}, \citenamefont {Hu}, \citenamefont
  {Ikeda}, \citenamefont {Szamel},\ and\ \citenamefont
  {Zamponi}}]{biroli2021unifying}%
  \BibitemOpen
  \bibfield  {author} {\bibinfo {author} {\bibfnamefont {G.}~\bibnamefont
  {Biroli}}, \bibinfo {author} {\bibfnamefont {P.}~\bibnamefont {Charbonneau}},
  \bibinfo {author} {\bibfnamefont {E.~I.}\ \bibnamefont {Corwin}}, \bibinfo
  {author} {\bibfnamefont {Y.}~\bibnamefont {Hu}}, \bibinfo {author}
  {\bibfnamefont {H.}~\bibnamefont {Ikeda}}, \bibinfo {author} {\bibfnamefont
  {G.}~\bibnamefont {Szamel}}, \ and\ \bibinfo {author} {\bibfnamefont
  {F.}~\bibnamefont {Zamponi}},\ }\href {\doibase 10.1103/PhysRevE.103.L030104}
  {\bibfield  {journal} {\bibinfo  {journal} {Phys. Rev. E}\ }\textbf {\bibinfo
  {volume} {103}},\ \bibinfo {pages} {L030104} (\bibinfo {year}
  {2021}{\natexlab{b}})}\BibitemShut {NoStop}%
\bibitem [{\citenamefont {Krzakala}\ and\ \citenamefont
  {Zdeborov{\'a}}(2009)}]{krzakala2009hiding}%
  \BibitemOpen
  \bibfield  {author} {\bibinfo {author} {\bibfnamefont {F.}~\bibnamefont
  {Krzakala}}\ and\ \bibinfo {author} {\bibfnamefont {L.}~\bibnamefont
  {Zdeborov{\'a}}},\ }\href {\doibase 10.1103/PhysRevLett.102.238701}
  {\bibfield  {journal} {\bibinfo  {journal} {Phys. Rev. Lett.}\ }\textbf
  {\bibinfo {volume} {102}},\ \bibinfo {pages} {238701} (\bibinfo {year}
  {2009})}\BibitemShut {NoStop}%
\bibitem [{\citenamefont {Coja-Oghlan}(2013)}]{coja2013quiet}%
  \BibitemOpen
  \bibfield  {author} {\bibinfo {author} {\bibfnamefont {A.}~\bibnamefont
  {Coja-Oghlan}},\ }\href
  {https://www.math.uni-frankfurt.de/~acoghlan/QuietPlanting.pdf} {\emph
  {\bibinfo {title} {Quiet Planting}}}\ (\bibinfo {year} {2013})\BibitemShut
  {NoStop}%
\bibitem [{\citenamefont {Charbonneau}\ \emph {et~al.}(2014)\citenamefont
  {Charbonneau}, \citenamefont {Jin}, \citenamefont {Parisi},\ and\
  \citenamefont {Zamponi}}]{CJPZ14}%
  \BibitemOpen
  \bibfield  {author} {\bibinfo {author} {\bibfnamefont {P.}~\bibnamefont
  {Charbonneau}}, \bibinfo {author} {\bibfnamefont {Y.}~\bibnamefont {Jin}},
  \bibinfo {author} {\bibfnamefont {G.}~\bibnamefont {Parisi}}, \ and\ \bibinfo
  {author} {\bibfnamefont {F.}~\bibnamefont {Zamponi}},\ }\href {\doibase
  10.1073/pnas.1417182111} {\bibfield  {journal} {\bibinfo  {journal} {Proc.
  Natl. Acad. Sci. U.S.A}\ }\textbf {\bibinfo {volume} {111}},\ \bibinfo
  {pages} {15025} (\bibinfo {year} {2014})}\BibitemShut {NoStop}%
\bibitem [{\citenamefont {Huang}\ \emph {et~al.}(2015)\citenamefont {Huang},
  \citenamefont {Wang},\ and\ \citenamefont {Yu}}]{huang2015non}%
  \BibitemOpen
  \bibfield  {author} {\bibinfo {author} {\bibfnamefont {Z.}~\bibnamefont
  {Huang}}, \bibinfo {author} {\bibfnamefont {G.}~\bibnamefont {Wang}}, \ and\
  \bibinfo {author} {\bibfnamefont {Z.}~\bibnamefont {Yu}},\ }\href@noop {}
  {\bibfield  {journal} {\bibinfo  {journal} {arXiv:1511.06672}\ } (\bibinfo
  {year} {2015})}\BibitemShut {NoStop}%
\bibitem [{\citenamefont {{Folena}}\ \emph {et~al.}(2022)\citenamefont
  {{Folena}}, \citenamefont {{Biroli}}, \citenamefont {{Charbonneau}},
  \citenamefont {{Hu}},\ and\ \citenamefont {{Zamponi}}}]{bcfhz2021long}%
  \BibitemOpen
  \bibfield  {author} {\bibinfo {author} {\bibfnamefont {G.}~\bibnamefont
  {{Folena}}}, \bibinfo {author} {\bibfnamefont {G.}~\bibnamefont {{Biroli}}},
  \bibinfo {author} {\bibfnamefont {P.}~\bibnamefont {{Charbonneau}}}, \bibinfo
  {author} {\bibfnamefont {Y.}~\bibnamefont {{Hu}}}, \ and\ \bibinfo {author}
  {\bibfnamefont {F.}~\bibnamefont {{Zamponi}}},\ }\href {\doibase
  10.48550/arXiv.2202.07560} {\  (\bibinfo {year} {2022}),\
  10.48550/arXiv.2202.07560}\BibitemShut {NoStop}%
\bibitem [{\citenamefont {Rizzo}\ and\ \citenamefont
  {Voigtmann}(2020)}]{rizzo2020solvable}%
  \BibitemOpen
  \bibfield  {author} {\bibinfo {author} {\bibfnamefont {T.}~\bibnamefont
  {Rizzo}}\ and\ \bibinfo {author} {\bibfnamefont {T.}~\bibnamefont
  {Voigtmann}},\ }\href {\doibase 10.1103/PhysRevLett.124.195501} {\bibfield
  {journal} {\bibinfo  {journal} {Phys. Rev. Lett.}\ }\textbf {\bibinfo
  {volume} {124}},\ \bibinfo {pages} {195501} (\bibinfo {year}
  {2020})}\BibitemShut {NoStop}%
\bibitem [{\citenamefont {Manacorda}\ \emph {et~al.}(2020)\citenamefont
  {Manacorda}, \citenamefont {Schehr},\ and\ \citenamefont
  {Zamponi}}]{manacorda2020numerical}%
  \BibitemOpen
  \bibfield  {author} {\bibinfo {author} {\bibfnamefont {A.}~\bibnamefont
  {Manacorda}}, \bibinfo {author} {\bibfnamefont {G.}~\bibnamefont {Schehr}}, \
  and\ \bibinfo {author} {\bibfnamefont {F.}~\bibnamefont {Zamponi}},\ }\href
  {\doibase 10.1063/5.0007036} {\bibfield  {journal} {\bibinfo  {journal} {J.
  Chem. Phys.}\ }\textbf {\bibinfo {volume} {152}},\ \bibinfo {pages} {164506}
  (\bibinfo {year} {2020})}\BibitemShut {NoStop}%
\bibitem [{foo()}]{footnote}%
  \BibitemOpen
  \href@noop {} {}\bibinfo {note} {Unlike $\chi^\mathrm{S}_4$, $\alpha_2$
  (Fig.~\ref{fig:a2}) does not vanish at short times, but is negative because
  the tracer velocity is fixed (with a random orientation) in the RLG instead
  of Gaussian distributed.}\BibitemShut {Stop}%
\bibitem [{\citenamefont {Charbonneau}\ \emph {et~al.}(2021)\citenamefont
  {Charbonneau}, \citenamefont {Charbonneau}, \citenamefont {Hu},\ and\
  \citenamefont {Yang}}]{charbonneau2021percolation}%
  \BibitemOpen
  \bibfield  {author} {\bibinfo {author} {\bibfnamefont {B.}~\bibnamefont
  {Charbonneau}}, \bibinfo {author} {\bibfnamefont {P.}~\bibnamefont
  {Charbonneau}}, \bibinfo {author} {\bibfnamefont {Y.}~\bibnamefont {Hu}}, \
  and\ \bibinfo {author} {\bibfnamefont {Z.}~\bibnamefont {Yang}},\ }\href
  {\doibase 10.1103/PhysRevE.104.024137} {\bibfield  {journal} {\bibinfo
  {journal} {Phys. Rev. E}\ }\textbf {\bibinfo {volume} {104}},\ \bibinfo
  {pages} {024137} (\bibinfo {year} {2021})}\BibitemShut {NoStop}%
\bibitem [{\citenamefont {Berthier}\ \emph {et~al.}(2016)\citenamefont
  {Berthier}, \citenamefont {Charbonneau}, \citenamefont {Jin}, \citenamefont
  {Parisi}, \citenamefont {Seoane},\ and\ \citenamefont {Zamponi}}]{BCJPSZ15}%
  \BibitemOpen
  \bibfield  {author} {\bibinfo {author} {\bibfnamefont {L.}~\bibnamefont
  {Berthier}}, \bibinfo {author} {\bibfnamefont {P.}~\bibnamefont
  {Charbonneau}}, \bibinfo {author} {\bibfnamefont {Y.}~\bibnamefont {Jin}},
  \bibinfo {author} {\bibfnamefont {G.}~\bibnamefont {Parisi}}, \bibinfo
  {author} {\bibfnamefont {B.}~\bibnamefont {Seoane}}, \ and\ \bibinfo {author}
  {\bibfnamefont {F.}~\bibnamefont {Zamponi}},\ }\href {\doibase
  10.1073/pnas.1607730113} {\bibfield  {journal} {\bibinfo  {journal} {Proc.
  Natl. Acad. Sci. U.S.A}\ }\textbf {\bibinfo {volume} {113}},\ \bibinfo
  {pages} {8397} (\bibinfo {year} {2016})}\BibitemShut {NoStop}%
\bibitem [{\citenamefont {Berthier}\ \emph {et~al.}(2020)\citenamefont
  {Berthier}, \citenamefont {Charbonneau},\ and\ \citenamefont
  {Kundu}}]{berthier2019finite}%
  \BibitemOpen
  \bibfield  {author} {\bibinfo {author} {\bibfnamefont {L.}~\bibnamefont
  {Berthier}}, \bibinfo {author} {\bibfnamefont {P.}~\bibnamefont
  {Charbonneau}}, \ and\ \bibinfo {author} {\bibfnamefont {J.}~\bibnamefont
  {Kundu}},\ }\href {\doibase 10.1103/PhysRevLett.125.108001} {\bibfield
  {journal} {\bibinfo  {journal} {Phys. Rev. Lett.}\ }\textbf {\bibinfo
  {volume} {125}},\ \bibinfo {pages} {108001} (\bibinfo {year}
  {2020})}\BibitemShut {NoStop}%
\bibitem [{\citenamefont {Rizzo}(2021)}]{rizzo2021path}%
  \BibitemOpen
  \bibfield  {author} {\bibinfo {author} {\bibfnamefont {T.}~\bibnamefont
  {Rizzo}},\ }\href {\doibase 10.1103/PhysRevB.104.094203} {\bibfield
  {journal} {\bibinfo  {journal} {Phys. Rev. B}\ }\textbf {\bibinfo {volume}
  {104}},\ \bibinfo {pages} {094203} (\bibinfo {year} {2021})}\BibitemShut
  {NoStop}%
\bibitem [{\citenamefont {Kirkpatrick}\ \emph {et~al.}(1989)\citenamefont
  {Kirkpatrick}, \citenamefont {Thirumalai},\ and\ \citenamefont
  {Wolynes}}]{KTW89}%
  \BibitemOpen
  \bibfield  {author} {\bibinfo {author} {\bibfnamefont {T.~R.}\ \bibnamefont
  {Kirkpatrick}}, \bibinfo {author} {\bibfnamefont {D.}~\bibnamefont
  {Thirumalai}}, \ and\ \bibinfo {author} {\bibfnamefont {P.~G.}\ \bibnamefont
  {Wolynes}},\ }\href {\doibase 10.1103/PhysRevA.40.1045} {\bibfield  {journal}
  {\bibinfo  {journal} {Physical Review A}\ }\textbf {\bibinfo {volume} {40}},\
  \bibinfo {pages} {1045} (\bibinfo {year} {1989})}\BibitemShut {NoStop}%
\bibitem [{\citenamefont {Keys}\ \emph {et~al.}(2011)\citenamefont {Keys},
  \citenamefont {Hedges}, \citenamefont {Garrahan}, \citenamefont {Glotzer},\
  and\ \citenamefont {Chandler}}]{KHGGC11}%
  \BibitemOpen
  \bibfield  {author} {\bibinfo {author} {\bibfnamefont {A.~S.}\ \bibnamefont
  {Keys}}, \bibinfo {author} {\bibfnamefont {L.~O.}\ \bibnamefont {Hedges}},
  \bibinfo {author} {\bibfnamefont {J.~P.}\ \bibnamefont {Garrahan}}, \bibinfo
  {author} {\bibfnamefont {S.~C.}\ \bibnamefont {Glotzer}}, \ and\ \bibinfo
  {author} {\bibfnamefont {D.}~\bibnamefont {Chandler}},\ }\href {\doibase
  10.1103/PhysRevX.1.021013} {\bibfield  {journal} {\bibinfo  {journal}
  {Physical Review X}\ }\textbf {\bibinfo {volume} {1}},\ \bibinfo {pages}
  {021013} (\bibinfo {year} {2011})}\BibitemShut {NoStop}%
\bibitem [{\citenamefont {Guiselin}\ \emph {et~al.}(2021)\citenamefont
  {Guiselin}, \citenamefont {Scalliet},\ and\ \citenamefont
  {Berthier}}]{guiselin2021microscopic}%
  \BibitemOpen
  \bibfield  {author} {\bibinfo {author} {\bibfnamefont {B.}~\bibnamefont
  {Guiselin}}, \bibinfo {author} {\bibfnamefont {C.}~\bibnamefont {Scalliet}},
  \ and\ \bibinfo {author} {\bibfnamefont {L.}~\bibnamefont {Berthier}},\
  }\href {\doibase 10.48550/arXiv.2103.01569} {\  (\bibinfo {year} {2021}),\
  10.48550/arXiv.2103.01569}\BibitemShut {NoStop}%
\end{thebibliography}%
\end{document}